\documentclass[conference]{IEEEtran}
\IEEEoverridecommandlockouts
\usepackage{amsthm}
\usepackage[hypertexnames=false]{hyperref}
\usepackage{orcidlink}
\usepackage{siunitx}
\usepackage[nolist]{acronym}
\usepackage{placeins}
\usepackage{multirow}
\usepackage{longtable}
\usepackage{float}
\usepackage{tabularx,booktabs, tabulary}
\usepackage{lipsum}
\usepackage{adjustbox}
\usepackage[normalem]{ulem}
\usepackage[style=ieee, sorting=nty, dashed=false, maxcitenames=1, mincitenames=1,isbn=false,url=true,doi=false,date=year]{biblatex}
\usepackage{amsmath,amssymb,amsfonts}
\usepackage[capitalise]{cleveref}
\Crefname{equation}{Eq.}{Eqs.}
\Crefname{figure}{Fig.}{Figs.}
\Crefname{tabular}{Tab.}{Tabs.}
\usepackage{algorithmic}
\usepackage{graphicx}
\usepackage{textcomp}
\usepackage{xcolor}
\newcommand{\BibTeX}{\textrm{B \kern -.05em \textsc{i \kern -.025em b} \kern -.08em
T \kern -.1667em \lower .7ex \hbox{E} \kern -.125emX}}

\newcommand{\swsys}{software architecture platform} % SW Architecture
\newcommand{\lowlayers}{platform, OS and drivers} % Lower Layers (BSW/MW)
\newcommand{\swcinterface}{data interface} %SW Data Interface
\newcommand{\samplesize}{nine}

\begin{document}

    \title{A Comparison of ROS 2 and AUTOSAR Adaptive Platform Against Industry-Elicited Automotive Middleware Requirements\\
    \thanks{This research is accomplished within the project ``\mbox{autotech.agil}'' (FKZ 01IS22088x). We acknowledge the financial support for the project by the Federal Ministry of Research, Technology and Space of Germany (BMFTR).}
    }
    \author{
        \IEEEauthorblockN{
            Lucas Hegerath\IEEEauthorrefmark{1}\IEEEauthorrefmark{3}\thanks{\IEEEauthorrefmark{3} Authors contributed equally to this paper.},
            David Philipp Klüner\IEEEauthorrefmark{1}\IEEEauthorrefmark{3},
            Philipp Pelcz\IEEEauthorrefmark{2}, \\
            Viswanatha Reddy Batchu\IEEEauthorrefmark{2},
            Marius Molz\IEEEauthorrefmark{1},
            Julius Kahle\IEEEauthorrefmark{1}, \\
            Thomas Schulik\IEEEauthorrefmark{2},
            Stefan Kowalewski\IEEEauthorrefmark{1},
            Alexandru Kampmann\IEEEauthorrefmark{1}
        }
        \IEEEauthorblockA{
            \IEEEauthorrefmark{1}Chair of Embedded Software\\
            RWTH Aachen University, Aachen, Germany\\
            Emails: \{hegerath, kluener, molz, kahle, kowalewski, kampmann\}@embedded.rwth-aachen.de
        }
        \IEEEauthorblockA{
            \IEEEauthorrefmark{2}ZF Group\\
            Friedrichshafen, Germany\\
            Emails: \{philipp.pelcz, viswanathareddy.batchu, thomas.schulik\}@zf.com
        }
    }
    \maketitle

    \begin{abstract}

        In software-defined vehicles, automotive middleware plays a fundamental role in enabling efficient communication, integration, and coordination among software components.
        This paper examines how well two of the currently most popular middleware frameworks, \ac{ROS 2} Jazzy and \acs{AUTOSAR} \acl{AP} R24-11, meet practical requirements elicited from automotive software engineers at one of the major automotive supplier companies, ZF Group.
        Our objective is to provide insight into an otherwise difficult-to-obtain industrial perspective and support a clearer understanding of priorities in the development and evaluation of middleware for automotive applications.

    \end{abstract}


    \section{Introduction}
    \label{sec:introduction}

    Modern vehicles increasingly rely on software to enable the wide range of functions they support and to meet evolving customer expectations~\cite{mckinsey_automotive_nodate, vetter_development_2020}.
    In this context, automotive middleware plays a critical role alongside emerging automotive E/E architectures~\cite{mckinsey_case_nodate, huck_next_2023}.
    These new architectures reduce vehicle wiring while increasing bandwidth, provide increased computing resources, and enable flexible software deployment.
    Middleware decouples automotive applications from the underlying hardware, offering essential services such as communication, execution management, and resource control across multiple \acp{ECU}~\cite{zhu_requirements-driven_2021}.
    At its core, middleware ensures interoperability, cooperation, and data exchange within the automotive software system.

    Academic research on middleware is well-established, often emphasizing flexibility, rapid development, and support for experimental innovation~\cite{kluner_modern_2024, henle_architecture_2022}.
    In contrast, industrial use of middleware imposes additional and sometimes divergent requirements, including robustness, security, modularity, and compliance with functional safety standards~\cite{autosar_explanation_sw_arch, spencer_how_2021}.
    Despite the growing importance of middleware in modern vehicles, publicly available, detailed insights and requirements from industrial practitioners remain limited.
    To the best of our knowledge, no prior work systematically reports industry-elicited requirements for automotive middleware, creating a gap between research focus and real-world automotive needs.

    To address this gap, we present middleware requirements collected from \samplesize{} software architects at ZF Group, a global Tier-1 supplier.
    Our objective is to provide an industrial perspective that can inform future middleware development and evaluation.
    In addition to reporting the elicited requirements, we analyze how \ac{ROS 2} Jazzy and \ac{AUTOSAR} \acl{AP} R24-11 satisfy these requirements, highlighting areas of alignment and divergence.

    Our contributions are as follows:
    \begin{itemize}
        \item A concise introduction to automotive middleware, highlighting its role in modern E/E architectures.
        \item A survey of requirements for automotive middlewares elicited from software architects at ZF Group.
        \item A comparative analysis of how two state-of-the-art middleware platforms, \ac{ROS 2} and \ac{AUTOSAR} \ac{AP}, satisfy the identified requirements.
    \end{itemize}

    \begin{figure}[tpb!]
        \centering
        \includegraphics[width=0.5\textwidth]{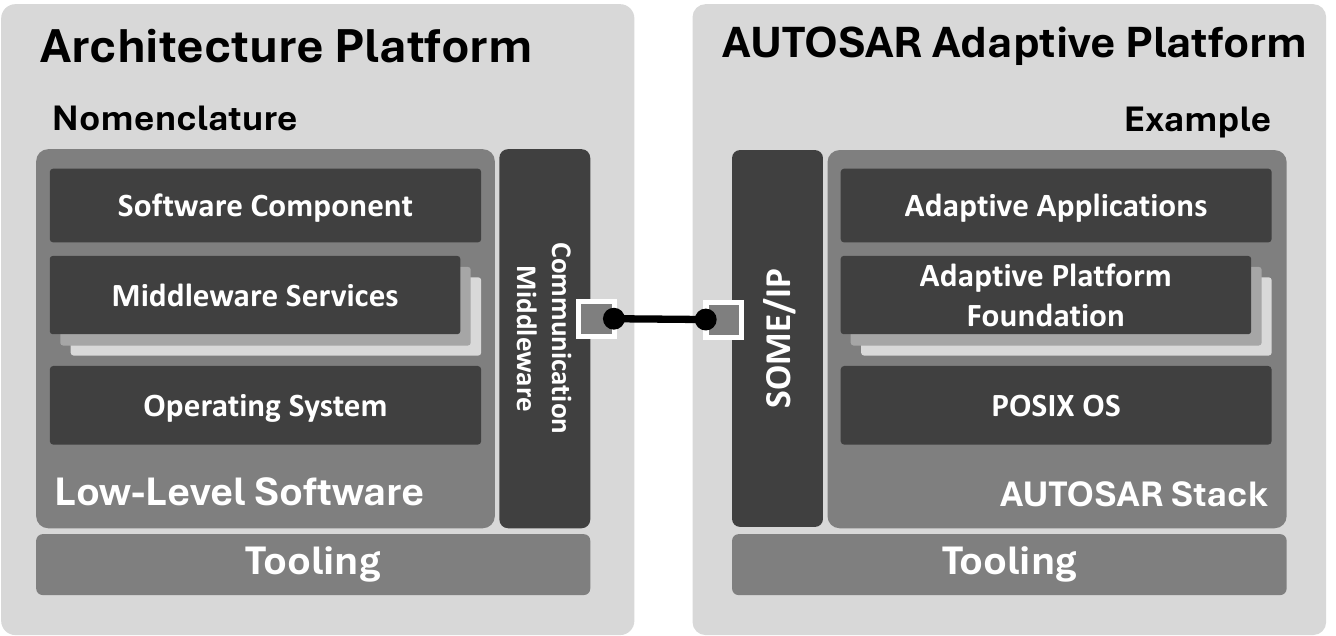}
        \caption{Illustration of current automotive software stacks and our nomenclature used. Within architecture platforms, software components use communication middlewares and middleware services, which in turn are built on the underlying \ac{OS}.}
        \label{fig:swstack}
    \end{figure}

    \section{Related Work}
    \label{sec:related_work}

    Beyond the requirements published by the \ac{AUTOSAR} Foundation, an academic discussion based on generic, platform-agnostic requirements for automotive middleware remains underexplored.
    To the best of our knowledge, we found no publications that directly report industry practitioners’ perspectives on middleware requirements.

    A partial view emerges from industry whitepapers, although these often remain high-level, emphasize topics aligned with corporate messaging, and provide only limited technical specificity.
    Many whitepapers concentrate on future vehicle architectures, particularly new \ac{E/E} architectures that consolidate computing resources on high-performance controllers while assigning I/O aggregation to zone controllers~\cite{huck_next_2023}.
    They frequently highlight software-related topics such as hypervisor-based virtualization and orchestration mechanisms.
    Whitepapers by \citeauthor{huck_next_2023}~\cite{huck_next_2023} and \citeauthor{aptiv_smart_2020}~\cite{aptiv_smart_2020}, for example, advocate containerization and Kubernetes-style orchestration for automotive \acp{ECU}, extended to meet hard real-time and safety requirements.
    To support function consolidation, they propose mixed-criticality execution environments with integrated functional safety mechanisms.
    Several sources further emphasize the need for middleware to isolate ASIL-classified components while enabling controlled interaction with QM software~\cite{spencer_how_2021, kristoferitsch_optimizing_2025}.
    Whitepapers also commonly discuss \ac{OTA} updates and continuous deployment.
    Because software evolves throughout the vehicle life cycle, update mechanisms should remain atomic, verifiable, and robust to failures~\cite{huck_next_2023}.
    In the area of communication, frequently referenced protocols include SOME/IP, DDS, MQTT, and REST, typically deployed via \ac{AUTOSAR} \ac{AP}~\cite{mayr_middleware_2020}.
    However, these sources generally articulate high-level goals rather than specific, implementable middleware requirements.

    Academic literature provides complementary perspectives on automotive middleware but largely reflects the research community’s thematic interests rather than industry-driven requirements. \citeauthor{henle_architecture_2022}~\cite{henle_architecture_2022} compare \ac{ROS 2} and \ac{AUTOSAR} \ac{AP}, using the latter as a baseline and evaluating \ac{ROS 2} as an open-source, community-driven alternative. \citeauthor{kukulicic_automotive_2022}~\cite{kukulicic_automotive_2022} present a survey of service-oriented automotive architectures, identifying a strong academic focus on security (addressed in 28\% of surveyed publications) and on safety (20\%). Other aspects relevant to middleware design, such as maintainability, integration, and adaptability, receive significantly less attention, appearing in only four of the reviewed studies.

    \section{Background}
    \label{sec:background}

    The \ac{E/E} architecture of a vehicle specifies the configuration and interconnection of its electrical and electronic hardware components.
    These include multiple \acp{ECU}, as well as sensors, power systems, wiring harnesses, and the \ac{IVN}~\cite{navale2015revolution}.
    Automotive \ac{E/E} architectures have undergone several paradigm shifts over recent decades~\cite{zhu_requirements-driven_2021}.

    Today, domain-based \ac{E/E} architectures remain the most widely deployed.
    In this architecture, each \ac{ECU} is dedicated to a specific function, and its computational resources are sized accordingly.
    These \acp{ECU} are typically microcontroller-based, and the \ac{IVN} topology is largely static, relying on multiple parallel fieldbus systems, such as \ac{CAN} and FlexRay, grouped into functional domains, e.g.\ powertrain or engine control~\cite{vetter_development_2020}.

    The transition toward software-defined vehicles increases the required flexibility and scalability of the architecture.
    Future systems must support continuous adaptation throughout the vehicle life cycle and handle rising demands in data throughput and software functionality, while preserving performance and safety guarantees~\cite{kluner_modern_2024}.

    Zone-based \ac{E/E} architectures have emerged as a response to these demands.
    Unlike domain-based architectures, they cluster \acp{ECU} according to their physical location.
    General-purpose \acp{ECU} are grouped into zones managed by higher-performance \acp{ZCU}.
    Communication within a zone is confined to its local \acp{ECU}, and only the \ac{ZCU} interface with other zones~\cite{brunner_automotive_2017}.
    These architectures primarily employ automotive Ethernet or combine it with legacy fieldbus technologies~\cite{wang_review_2024}.
    While zone-based designs simplify wiring and reduce the number of discrete fieldbuses, they also introduce new complexity in system management.

    To address this complexity, automotive middlewares provide an abstraction layer on top of the \ac{E/E} architecture.
    Beyond enabling application-to-application communication, they support execution management, monitoring, configuration, error handling, and update mechanisms.
    \cref{fig:swstack} relates our nomenclature on the left side with a representative middleware-based software stack for zone-based \ac{E/E} architectures on the right side, showing how software components interact with communication middlewares and middleware services, which themselves rely on the underlying \ac{OS}.

    Two prominent middleware frameworks are \ac{ROS 2} and \ac{AUTOSAR} \ac{AP}.
    While the \ac{AUTOSAR} Foundation, an international consortium of automotive \acp{OEM} and suppliers, governs \ac{AUTOSAR} \ac{AP}, \ac{ROS 2} originates from the \ac{CPS} and robotics community as an open-source project~\cite{henle_architecture_2022}.
    Both provide layered architectures that abstract applications from the underlying \ac{OS} and hardware.

    In \ac{AUTOSAR} \ac{AP}, most middleware-related concerns are contained within the modular layer \ac{ARA}.
    The \aclp{FC} within \ac{ARA} offer \acp{API} for communication, execution management, health monitoring, configuration, and various security and safety mechanisms.

    \ac{ROS 2} implements a multi-layered architecture with different levels of abstraction.
    Its two key interfaces are the client library (\textit{rcl}) and the middleware interface (\textit{rmw}). The \textit{rmw} layer exposes the \ac{API} between the communication protocols implemented in the \ac{OS} and the \ac{ROS 2} software stack.
    The higher-level \textit{rcl} layer manages tasks such as interacting with \textit{rmw}, orchestrating application execution, and handling errors.
    In contrast to the standardized \ac{AUTOSAR} \ac{AP} approach, the open-source nature of \ac{ROS 2} prioritizes flexibility over strict conformity to industry specifications~\cite{henle_architecture_2022}.

    \section{Study Design and Scope}
    \label{sec:study_design}

    We aim to provide a structured perspective on the requirements that industry practitioners place on automotive middleware.
    To do so, we interviewed \samplesize{} of ZF Group's automotive software architects and documented the resulting requirements as they were expressed.

    We conducted the elicitation within the Automotive SPICE framework, a process assessment model based on ISO/IEC~15504 and widely adopted in the automotive industry~\cite{vda_working_group_13_automotive_2023} to ensure consistent, high-quality software and system development.
    Automotive SPICE defines a set of process areas and capability levels, including SYS.2 (Requirements Elicitation), which covers the systematic collection, documentation, and consolidation of stakeholder needs.
    By conducting our study within this process area, we ensured that the collected requirements follow established industrial practices.
    Within this SPICE-compliant context, the only constraint placed on the participants was that the requirements must not preempt or constrain later architectural decisions in the development process, e.g.\ explicitly require a service-oriented architecture (SOA).

\begin{figure}[tbp!]
        \centering
        \includegraphics[width=0.5\textwidth]{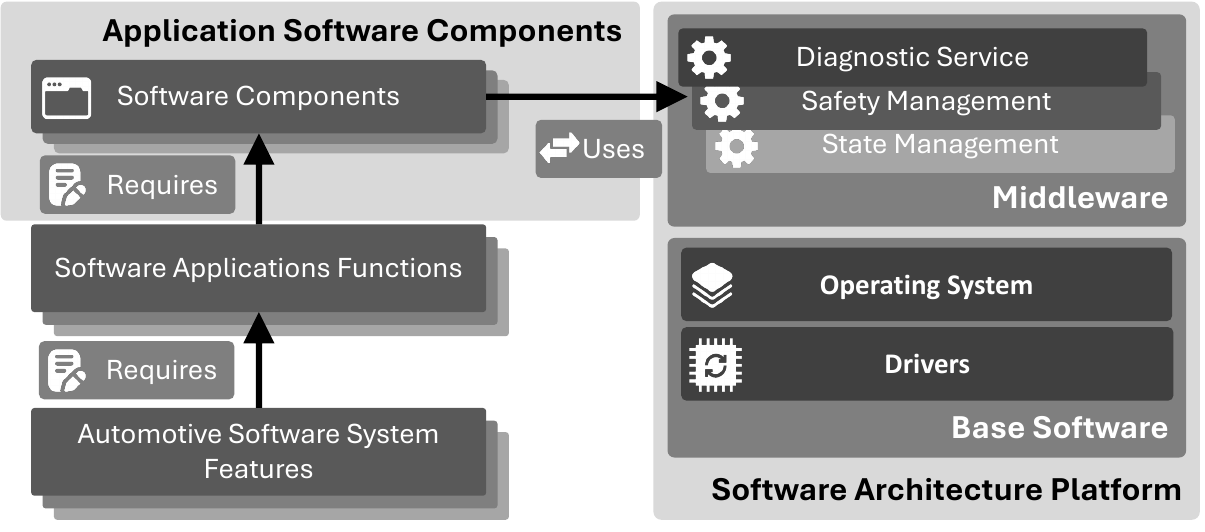}
        \caption{Illustration of the interaction of our terms. Automotive software system features are provided by multiple application functions, each of which is implemented by individual software components. These components rely on the software architecture platform for specific services.}
        \label{fig:nomenclature}
    \end{figure}

    We adopt an architecture-agnostic nomenclature and dependency model, illustrated in \cref{fig:nomenclature}.
    Automotive software system features are realized by multiple software application functions, each of which is implemented by individual software components.
    These software components depend on the underlying software architecture platform, which provides essential services through its middleware, operating system, and base software layers.

    The scope of our study focuses on the non-functional requirements that application-layer software components place on the middleware infrastructure.
    We intentionally exclude broader system-level aspects such as \ac{OTA} to maintain a clear focus on middleware concerns.

    Beyond eliciting requirements, we also assess how well existing middleware platforms fulfill them.
    For this purpose, we analyze \ac{AUTOSAR} \acl{AP} and \ac{ROS 2} using their official specifications and documentation.
    Based on this evaluation, we derived implications for the further development of emerging middleware platforms to align with the identified industry demands.

    We acknowledge that the single-company scope limits the generalizability of our findings.
    As with similar surveys, potential threats to validity include participant selection bias, organizational influences on responses, and limited replicability.
    Nevertheless, as discussed in \cref{sec:related_work}, our results provide concrete, practice-based insights that are otherwise difficult to obtain and that can inform future middleware development and evaluation.

    \section{Requirements on Automotive Middlewares}
    \label{sec:specifications}

    This section presents the results of our survey, summarized in \cref{tab:middleware_comparison}.
    The survey yielded 33 requirements, grouped into nine categories, including communication, execution control, error handling, and system state management.

    Although the requirements were formulated independently of any specific programming language, a noticeable bias toward C/C++ emerged, reflecting the predominant tooling and practices within the company.
    Specifically, the requirements assume components with explicit init/deinit routines, periodic or event-driven runnables, strongly typed data interfaces, buffered or immediate data access, service/method calls, fault reporting with fallback values, configurable execution behavior, and parameter read/write mechanisms, which resemble patterns in C/C++ embedded component design.

    We note that participants did not strictly separate middleware and operating system responsibilities, as some middleware implementations may delegate lower-level tasks, such as thread scheduling, to the operating system.

    \section{Comparison with AUTOSAR AP and ROS 2}
    \label{sec:requirement_fulfillment}

    This section compares the collected requirements with \ac{AUTOSAR} \ac{AP} and \ac{ROS 2}, which are frequently discussed for automotive middleware applications~\cite{henle_architecture_2022}.
    ZF Group members conducted the comparison with \ac{ROS 2}, while authors from RWTH Aachen compared the requirements against the \ac{AUTOSAR} \ac{AP} standard.
    \cref{tab:middleware_comparison} lists the requirements and the corresponding comparison results.
    Overall, \ac{ROS 2} fulfills 13 requirements, partially fulfills 1, and does not fulfill 19, whereas \ac{AUTOSAR} \ac{AP} fulfills 23 requirements, partially fulfills 1, and does not fulfill 9.

    \clearpage
    \onecolumn
    {\footnotesize
    \renewcommand{\arraystretch}{1.16}
        \begin{longtable}{p{0.7cm} p{7cm} p{4.0cm} p{5cm}}
            \caption{Comparison between the requirements and ROS 2 / AUTOSAR AP} \label{tab:middleware_comparison} \\
            \toprule
            Name & Specification & Robot Operating System 2 & \ac{AUTOSAR} Adaptive Platform \\
            \midrule
            \endfirsthead
            \caption[]{Comparison between the requirements and ROS 2 / AUTOSAR AP (continued)} \\
            \toprule
            Name & Specification & Robot Operating System 2 & \ac{AUTOSAR} Adaptive Platform \\
            \midrule
            \endhead
            \multicolumn{4}{l}{\textbf{Data Types}}\\
            DR 1 & The \swsys{} shall support the following datatypes for data interfaces to \lowlayers{} or \ac{SWC}: uint[8,16,32,64], int[8,16,32,64], float[32,64], array, enum, struct, union, boolean, char. & \checkmark{} Full support, uses the \ac{OMG} \ac{IDL} standard \cite{macenski_robot_2022} &  \checkmark{} Supported in the \ac{OSI} and COM \acp{FC} \cite{autosar_osi, autosar_com_management}. \\
            DR 2 & The \swsys{} shall support data interface units, scaling, offset configuration (incoming or outgoing data units) and data range configuration for data validation checks. & \texttimes{} Not supported by \ac{ROS 2} \cite{object_management_group_interface_nodate,macenski_robot_2022}. & $\circ{}$ Units can be specified via the \ac{ARXML} configuration, but conversion needs to be implemented manually \cite{autosar_com_management}. \\
            DR 3 & The \swsys{} shall support timeout configuration of the \swcinterface{} coming from other \ac{SWC} & \checkmark{} Supported only for Request-response communication \cite{open_robotics_quality_2025}. & \checkmark{} Supported via the COM \ac{FC} \cite{autosar_com_management}. \\
            DR 4 & The \swsys{} shall support the immediate or buffered access of the \swcinterface. & \checkmark{} Supported using \ac{ROS 2} \ac{QoS} \ac{AP}I \cite{open_robotics_quality_2025} & \checkmark{} Supported via the COM and RDS \acp{FC} \cite{autosar_com_management, autosar_rds}. \\
            DR 5 & The \swsys{} shall support a service/method interface for \acp{SWC} with which other \acp{SWC} can call a service/method at will. & \checkmark{} Supported using \ac{ROS 2} Services and Actions \cite{open_robotics_quality_2025}. & \checkmark{} Supported via the COM \ac{FC} \cite{autosar_com_management}. \\
            \midrule
            \multicolumn{4}{l}{\textbf{Scheduling}}\\
            SR 1 & The \swsys{} shall support running of \acp{SWC} initialization at startup, and deinitialization at shutdown, independent of its main algorithm. & \checkmark{} Supported, either via a constructor or the lifecycle state \textit{Unconfigured}\cite{macenski_robot_2022} & \checkmark{} Supported. The \ac{EM} handles the lifecycle of applications, including initialization and deinitialization phases \cite{autosar_exec_management}. \\
            SR 2 & The \swsys{} shall support the cyclic execution of the \acp{SWC} function with a defined cycle time. & \checkmark{} Supported. \ac{ROS 2} implements timer-triggered (periodic) callbacks\cite{macenski_robot_2022} &  \texttimes{} Not supported, custom implementation with \ac{POSIX} timers via \ac{OSI} possible \cite{autosar_osi}. \\
            SR 3 & The \swsys{} shall support configuration of multiple cyclic runnables with different cycle times within \acp{SWC}. & \checkmark{} Parallel execution using a multi-threaded executor is possible \cite{macenski_robot_2022} & \texttimes{} Not supported, custom implementation with \ac{POSIX} timers via \ac{OSI} possible \cite{autosar_osi}. \\
            SR 4 & The \swsys{} shall support the sequencing/calling order of the \acp{SWC} runnables. & \checkmark{} Callback execution in order of callback registration at runtime (\texttt{rclcpp}) \cite{yang_exploring_2020} or user-specified (\texttt{rclc}) \cite{staschulat_rclc_2020} & \texttimes{} Not supported, implementation via application logic necessary \cite{autosar_exec_management, autosar_osi}. \\
            SR 5 & The \swsys{} shall support the \ac{SWC} runnable execution on interface data reception, error status changes, and mode changes & \checkmark{} Supported through \texttt{rclc} \cite{staschulat_rclc_2020, macenski_robot_2022} & \checkmark{} Supported through service-oriented architecture \cite{autosar_com_management}. \\
            SR 6 & The \swsys{} shall support execution offsets for \ac{SWC} runnables. & \texttimes{} Not supported, can be manually implemented using \texttt{rclc} \cite{staschulat_rclc_2020} & \texttimes{} Not supported, custom implementation with \ac{OSI} possible \cite{autosar_osi}. \\
            SR 7 & The \swsys{} shall support preemptive/non-preemptive scheduling of the \ac{SWC} runnables. & \texttimes{} Framework only provides non-preemptive executors\cite{teper_timing-aware_2023}  & \texttimes{} Not supported on application level. Configuration on \ac{OS} level possible with \ac{POSIX} scheduling policy via \ac{OSI} \cite{autosar_osi}. \\
            SR 8 & The \swsys{} shall support a debounce time within repeated \ac{SWC} runnable execution. & \texttimes{} Not supported. & \texttimes{} Not supported, custom implementation with \ac{OSI} possible \cite{autosar_osi}. \\
            SR 9 & The \swsys{} shall support configuration of function runnable priority for scheduling. & \texttimes{} Not supported. No native Priority-based Executor available, but parallel execution of \textit{callback groups} available\cite{open_robotics_executors_2024}. & \texttimes{} Not supported on application level. Configuration on \ac{OS} level possible with \ac{POSIX} priorities via \ac{OSI} \cite{autosar_osi}. \\
            \midrule
            \multicolumn{4}{l}{\textbf{Data Queues}}\\
            DQR 1 & The \swsys{} shall support fixed length queued \swcinterface{}, with the most recent data at the end of the queue. & \checkmark{} Supported by communication middleware via \ac{QoS} configuration \cite{open_robotics_quality_2025}. & \checkmark{} Supported via the COM \ac{FC} \cite{autosar_com_management}. \\
            \midrule
            \multicolumn{4}{l}{\textbf{Parametrization}}\\
            PR 1 & The \swsys{} shall support reading parameters of the \ac{SWC} during startup. & \checkmark{} Supported using \ac{ROS 2} parameter \ac{API} \cite{open_robotics_monitoring_2025, macenski_robot_2022}. Parameters can be retrieved by the application and used.  & \checkmark{} Supported with the Persistency \ac{FC}. \ac{AP} Applications can read startup parameters from Persistency storage. The manifest describes the key-value or file storages, and the \ac{AP}I allows retrieval \cite{autosar_per}. \\
            PR 2 & The \swsys{} shall support writing/learning parameters during run time or shutdown. & \texttimes{} Not supported, node parameters are not persistent and lost at restart \cite{open_robotics_parameters_nodate}. & \checkmark{} Supported with the Persistency \ac{FC}, parameters are modifiable at runtime or shutdown. Configurable storage allows read/write access and maintains persistence through restarts \cite{autosar_per}. \\
            \midrule
            \multicolumn{4}{l}{\textbf{Error Handling}}\\
            EH 1 & The \swsys{} shall support reporting of functional/non-functional, error/fault status to common error/fault manager. & \checkmark{} Supported, custom errors can be raised and logged \cite{macenski_robot_2022,henle_architecture_2022} & \checkmark{} The \acl{AP} provides mechanisms for \acp{SWC} to report both functional and non-functional errors through the Diagnostic Management (DM) \ac{FC} \cite{autosar_diag}. \\
            EH 2 & The \swsys{} shall support reading the fault status from fault manager to \ac{SWC}. & \texttimes{} Not supported. \ac{ROS 2} implements no central fault manager. & \checkmark{} \acp{SWC} can query the status of diagnostic events, including their current state (e.g., active, inactive), through the Diagnostic Management \acp{AP}I \cite{autosar_diag}. \\
            EH 3 & The \swsys{} shall support reporting parameter error/fault status, and shall substitute default/invalid values. & \checkmark{} Supported, as \ac{ROS 2} allows for default parameter values \cite{open_robotics_monitoring_2025}.  & \texttimes{} Not supported by \ac{AUTOSAR} \ac{AP}, implementation on application level possible \cite{autosar_diag}. \\
            EH 4 & The \swsys{} shall support configuration of reporting \swcinterface{}, communication errors/faults status (e.g., missing/dropout, data integrity, sequencing and cryptographic checks, etc.), and shall substitute default/invalid values to data interfaces. & $\circ{}$ Partially supported, the security related functions are present in the \ac{ROS 2} security package \cite{open_robotics_ros_2025}. & \checkmark{} Supported, as \ac{SWC} can monitor communication statuses and implement fallback strategies, including default values in case of detected communication errors \cite{autosar_diag, autosar_com_management}. \\
            EH 5 & The \swsys{} shall support error/fault maturation times or maturation count limits. & \texttimes{} Not supported. \ac{ROS 2} lacks a central error/fault manager. & \checkmark{} Diagnostic Management (DM) enables debouncing algorithms to filter transient faults using maturation times or counts before an error is considered active \cite{autosar_diag}. \\
            EH 6 & The \swsys{} shall support the dependent error/fault configuration of subsequent errors/faults which are active due to common cause faults. & \texttimes{} Not supported. \ac{ROS 2} lacks a central error/fault manager. & \checkmark{} \ac{AP} supports dependent or related diagnostic events, letting applications link events so that the activations can affect another \cite{autosar_diag}. \\
            \midrule
            \multicolumn{4}{l}{\textbf{Functional Safety}}\\
            FSR 1 & The \swsys{} shall support safety reactions like \ac{SWC} degradation or inhibition, including actions to reach fail-safe based on single or multiple error/fault combination. & \texttimes{} Not supported. \ac{ROS 2} implements no centralized safety reaction mechanism.  & \checkmark{}Supported with \ac{PHM} and \ac{EM} \acp{FC}, these identify faults and switch to fail-safe or fail-operational states upon errors. Configurations are handled via \ac{ARXML} \cite{autosar_exec_management, autosar_phm}. \\
            \midrule
            \multicolumn{4}{l}{\textbf{Operating Modes}}\\
            MR 1 &  The \swsys{} shall support mode change request to overall \swsys{} from \ac{SWC}. & \texttimes{} Not supported in \ac{ROS 2}, but present in micro-ROS \cite{eprosima_introduction_2024}. \ac{ROS 2} implements no central mode management. & \texttimes{} Not supported by \ac{AUTOSAR} \ac{AP}. \\
            MR 2 &  The \swsys{} shall support \ac{SWC} reading overall \swsys{} mode. & \texttimes{} Not supported in \ac{ROS 2}, but present in micro-ROS \cite{eprosima_introduction_2024}. \ac{ROS 2} implements no central mode management. & \checkmark{} \acp{SWC} can query the current operational mode by accessing the State Management service \cite{autosar_sm}. \\
            MR 3 &The \swsys{} shall support \ac{SWC} execution/shutdown based on mode change (e.g., diagnostic mode). & \texttimes{} Not supported in \ac{ROS 2}, but present in micro-ROS \cite{eprosima_introduction_2024}. \ac{ROS 2} implements no central mode management. & \checkmark{} Supported by \ac{EM} \ac{FC}, \acp{SWC} can be grouped into Functional Groups, linked to specific modes. Functional Groups enable the activation or deactivation of \acp{SWC} in response to mode transitions \cite{autosar_exec_management}. \\
            \midrule
            \multicolumn{4}{l}{\textbf{Allocation}}\\
            AR1 & The \swsys{} shall support configuration of \ac{SWC} runnable stack size allocation. & \texttimes{} Not supported, due to \ac{ROS 2} unbounded length types and focus on \ac{POSIX} platforms. & \checkmark{} Stack size can be configured using \ac{ARXML} and \ac{EM} \ac{FC}, enabling resource allocation per \ac{SWC} \cite{autosar_exec_management}.  \\
            AR2 & The \swsys{} shall support \ac{SWC} configuration to ensure reliable execution (e.g \ac{SWC} running on Checker Core) of \ac{SWC} runnable & \texttimes{} Not supported, \ac{ROS 2} by default operates on \ac{POSIX} platforms and has no such concepts. & \checkmark{} Core Affinity and other hardware configuration is supported via \ac{ARXML} configurations and the \ac{EM} \ac{FC} \cite{autosar_exec_management}. \\
            \midrule
            \textbf{Watchdog}  \\
            WD1 & The \swsys{} shall support monitoring of the aliveness of \ac{SWC} runnables running periodically with fixed cycle times. & \texttimes{} Not supported, however some limited information about communication activity is accessible using the diagnostics topic. & \checkmark{} Supports Alive Supervision through the \ac{PHM} \ac{FC}, which checks if entities run within expected time frames \cite{autosar_phm}. \\
            WD2 & The \swsys{} shall support monitoring of the execution time limits of \ac{SWC} runnables. & \texttimes{} Not supported, \ac{ROS 2} implements no such concept. & \checkmark{} Supports Deadline Supervision via \ac{PHM}, ensuring that the execution time between specified checkpoints in a controlled entity meets defined time constraints \cite{autosar_phm}. \\
            WD3 & The \swsys{} shall support monitoring of the \acp{SWC} runnables order of execution. & \texttimes{} Not supported. Introspection to this extent is not implemented in \ac{ROS 2}. & \checkmark{} Supports Logical Supervision in the \ac{PHM} \ac{FC} to monitor the sequence of execution of checkpoints within supervised entities \cite{autosar_phm}. \\
            WD4 & The \swsys{} shall support monitoring of \acp{SWC} runnables of a function with all possible combinations of alive deadline and logical supervision. & \texttimes{} Not supported. Due to missing monitoring concepts. & \checkmark{} Supports the configuration of various supervision types, such as Alive, Deadline, and Logical, individually or in combination, for each supervised entity in the \ac{PHM} \cite{autosar_phm}.\\
            \bottomrule
        \end{longtable}
        \begin{center}
            The \checkmark{} icon indicates full support for a requirement, $\circ$ indicates partial support for a given requirement, while the \texttimes{} icon shows no support for a requirement.
        \end{center}
        \par}
    \clearpage
    \twocolumn

    The comparison is intentionally unweighted and reports requirement coverage, not implementation effort.
    We observed that requirements differ in implementation complexity: some can be added with application-level libraries, while others require deeper integration with middleware internals or the operating system interface.

    \subsection{AUTOSAR AP}
    \label{subsec:autosar}

    \ac{AUTOSAR} \ac{AP} fulfills a broad set of requirements, particularly in data type support and service-oriented communication.
    The platform provides well-defined interfaces, serialization, and reliable communication mechanisms suitable for distributed applications, thereby meeting most of the collected requirements.
    For safety and fault tolerance, dedicated functional clusters provide extensive capabilities.

    However, the current \ac{AUTOSAR} \ac{AP} standard does not meet the requirement for precise execution control~\cite{autosar_timing_extension}, and communication and execution remain non-deterministic, although recent work proposes mechanisms to improve determinism~\cite{9116430, BELLASSAI2025103390}.
    Earlier versions of this standard included a deterministic client within the execution management cluster, but this feature was omitted in recent releases, potentially in favor of the SL LET Timing Extension for deterministic scheduling and communication~\cite{autosar_timing_extension}.
    Furthermore, while \ac{AUTOSAR} \ac{AP} fulfills two of the scheduling-related requirements, the remaining scheduling capabilities must be implemented by the user via the \ac{OSI}.

    Overall, \ac{AUTOSAR} \ac{AP} fulfills the majority of requirements, with the main limitation being advanced execution control.

    \subsection{ROS 2}
    \label{subsec:ros2}

    \ac{ROS 2} meets many communication and data type requirements through its use of DDS and \ac{IDL}-based message definitions.
    Its executor-based execution model allows in-process scheduling of callbacks, messages, and timers, though it remains constrained by the underlying operating system’s scheduling~\cite{choi_2021}.
    Basic configuration and diagnostics offer partial support for parameterization and error handling.

    However, \ac{ROS 2} lacks several concepts essential for series-production automotive systems.
    Notably, it does not provide operation modes, allocation mechanisms, or watchdog supervision, and therefore does not fulfill any requirements in these categories.
    It also does not offer built-in failover strategies or deployment-oriented safety features, as \ac{AUTOSAR} \ac{AP} does with the \ac{PHM} \ac{FC}, which are typically required in automotive environments.
    Furthermore, \ac{ROS 2} has issues regarding real-time capability for safety-critical applications~\cite{teper_timing-aware_2023, casini_response-time_2019}.
    These findings align with prior evaluations concluding that \ac{ROS 2} prioritizes ease of development and flexibility over deployment-oriented robustness~\cite{kluner_modern_2024}.

    In summary, \ac{ROS 2} satisfies a substantial subset of communication-related and structural requirements but lacks capabilities needed for safety-critical or production-grade deployments.
    For deterministic scheduling and communication in \ac{AUTOSAR} \ac{AP}, the SL LET Timing Extension can be employed~\cite{autosar_timing_extension}.
    For deploying \ac{ROS 2} in safety-critical systems, industrial forks such as the Apex.AI variant of \ac{ROS 2} may be considered.
    Across both technologies, the scheduling category displays the highest number of unfulfilled requirements.
    This emphasizes the need for future research into deterministic, analyzable, and verifiable scheduling mechanisms for distributed automotive software systems.

    \subsection{Framework Tradeoffs}
    \label{subsec:tradeoffs}

    The results indicate different design priorities.
    \ac{ROS 2} prioritizes openness, development speed, and a broad ecosystem, which helps with rapid feature development and integration.
    This comes with tradeoffs for production deployment in automotive systems, where standardized supervision, persistent parameter handling, and built-in safety management are expected.

    \ac{AUTOSAR} \ac{AP} prioritizes standardized diagnostics, safety-related services, and deployment control through dedicated functional clusters.
    These capabilities improve readiness for series production, but advanced runnable-level scheduling and timing control are still limited in the base platform and often delegated to \ac{OSI}-based integration.

    \section{Conclusion}
    \label{sec:conclusion}

    In this study, we collected, structured, and analyzed practitioner-defined requirements for automotive middleware based on input from software architects at ZF Group.
    We identified 33 requirements across nine categories, covering communication, data modeling, scheduling, and safety-related features such as watchdog mechanisms.
    The responses showed that participants were strongly familiar with middleware frameworks currently used in production, particularly \ac{AUTOSAR} Classic.

    We evaluated how well \ac{ROS 2} and \ac{AUTOSAR} \ac{AP} fulfill the elicited requirements by analyzing their official specifications and documentation.
    The comparison indicates that \ac{AUTOSAR} \ac{AP} satisfies a broader range of requirements. \ac{ROS 2} aligns with its prototyping-oriented design philosophy but shows notable gaps, especially in safety-related capabilities and features required for series production.

    To improve the generalizability of our results, we plan to extend the survey to additional companies and broaden the empirical basis of future analyses.

    \section{Acknowledgement}
    \label{sec:acknowledgement}

    This research is accomplished within the project ``\mbox{autotech.agil}'' (FKZ 01IS22088x).
    We acknowledge the financial support for the project by the Federal Ministry of Research, Technology and Space of Germany (BMFTR).

    \begin{acronym}[ROS 2]
    \acro{ARXML}{AUTOSAR XML}
    \acro{OMG}{Object Management Group}
    \acro{EM}{Execution Management}
    \acro{OSI}{Operating System Interface}
    \acro{IDL}{Interface Definition Language}
    \acro{POSIX}{Portable Operating System Interface}
    \acro{PHM}{Platform Health Management}
    \acro{FC}{Functional Cluster}
    \acro{ECU}{Electronic Control Unit}
    \acro{AP}{Adaptive Platform}
    \acro{OS}{Operating System}
    \acro{ROS 2}{Robot Operating System 2}
    \acro{OTA}{Over-the-air}
    \acro{V2X}{Vehicle-to-Everything}
    \acro{E/E}{Electrical/Electronic}
    \acro{IVN}{In-Vehicle Network}
    \acro{CAN}{Controller Area Network}
    \acro{ZCU}{Zone Control Unit}
    \acro{AUTOSAR}{Automotive Open System Architecture}
    \acro{OEM}{Original Equipment Manufacturer}
    \acro{CPS}{Cyber-Physical Systems}
    \acro{ARA}{AUTOSAR Runtime Environment for Adaptive Applications}
    \acro{API}{Application Programming Interface}
    \acro{SWC}{Software Component}
    \acro{QoS}{Quality of Service}
    \acro{AV}{Autonomous Vehicle}
    \end{acronym}

    \printbibliography

\end{document}